\documentclass[fleqn,twoside]{article}
\usepackage{espcrc2}

\usepackage{graphicx}
\usepackage[figuresright]{rotating}
\usepackage{epsfig}

\newcommand{\AmS}{{\protect\the\textfont2
  A\kern-.1667em\lower.5ex\hbox{M}\kern-.125emS}}

\title{Mesonic Spectrum from a Dynamical Gravity/Gauge model}

\author{W. de Paula\address[]{Dep. de F\'\i sica, Instituto Tecnol\'ogico de Aeron\'autica,
CTA, 12228-900. S\~ao Jos\'e dos Campos, Brazil}%
        \thanks{Supported by CAPES} and
        T. Frederico\addressmark.}

\begin{document}

\begin{abstract}
Within a formulation of a Dynamical AdS/QCD model we calculate the
spectrum of light flavored mesons. The background fields of the
model correspond to an IR deformed Anti de Sitter metric coupled
to a dilaton field. Confinement comes as a consequence of the
dilaton dynamics coupled to gravity. Additionally to the
Regge-like spectrum of light- scalar, vector and higher spin
mesons, we obtain the decay width of  scalar mesons into two
pions. \vspace{1pc}
\end{abstract}

\maketitle

\section{Introduction}

The non perturbative aspects of strong interaction are extremely
difficult to treat analytically. Most of our knowledge about the
low energy limit of the strong force comes from Lattice
calculations. In this context, the AdS/CFT correspondence
\cite{Maldaconj} represents an attractive alternative to
investigate non perturbative aspects of gauge theories. This
duality connects the string theory amplitudes on asymptotically
AdS $\times X$ space-time into gauge invariant, local operators of
a conformal field theory (CFT). As a consequence, the notoriously
complex strong-coupling regime of large-$N_{c}$ gauge theories can
be approximated (in low-curvature regions) by weakly coupled and
hence analytically treatable classical gravities.

Extensions of this idea to Quantum Chromodynamics (QCD) either
start from specific D-brane setups in ten- (or five-) dimensional
supergravity \cite{Klebanov:2000nc,Klebanov:2000hb,Maldacena:2000yy,Erdmenger:2007cm,waynebianchi}
and derive the corresponding gauge theory properties,
or try to guess a suitable background and to improve it in
bottom-up fashion by comparing the predictions to QCD data. The
first bottom-up model (Hard Wall) developed by Polchinsky and
Strassler \cite{Polchinski} shows that the conformal invariance of
AdS$_{5}$ in the UV limit implements the counting rules which
govern the scaling behavior of hard QCD scattering amplitudes. An
infrared cutoff on the fifth dimension at the QCD scale $\Lambda
_{QCD}$ gives the mass gap and a discrete hadron spectrum. This
model reproduces a huge amount of hadron phenomenology\cite{hwph}.
On the other hand it does not reproduce  Regge trajectories on the
mass spectrum ($M^2 \times n$). To correct this shortcoming Karch,
Katz,  Son and Stephanov developed the Soft-Wall
model\cite{Karch}. In this approach the AdS$_{5}$ geometry is kept
intact while an additional dilaton background field is responsible
for the conformal symmetry breaking. This dilaton soft-wall model
indeed generates linear Regge trajectories $m_{n,S}^{2}\sim n+S$
for light-flavor mesons of spin $S$ and radial excitation $n$.
(Regge behavior can alternatively be encoded via IR deformations
of the AdS$_{5}$ metric \cite{Kruczenski,FBT}.)

However, the resulting vacuum expectation value (vev) of the
Wilson loop in the dilaton soft wall model does not exhibit the
area-law behavior which a linearly confining static
quark-antiquark potential would generate. It happens because the
model uses an AdS metric which is not of a confining type by the
Wilson Loop analysis \cite{Malda2,Rey01}. In addition the soft
wall model background is not a solution of a dual classical
gravity theory. Therefore, one has to impose all gauge theory
vacuum properties (confinement, chiral symmetry breaking and
condensates) in an ad-hoc manner, and the desired connection to
the dynamics of a QCD dual remains untouched \cite{for07}.

Csaki and Reece \cite{Csaki} analyzed the solutions of a 5d
dilaton-gravity Einstein equations (see also\cite{Kiritsis}) using
the formalism of superpotential. Their conclusion is that it would
not be possible to solve those equations and obtain a linear
confining background. They also suggest that it would be possible
to get a solution by analyzing a tachyon-dilaton-graviton model.
This idea was successfully implemented by Batell and Gherghetta
\cite{Batell}.

We took an alternative route and we showed\cite{dePaulaPRD09} that
a linear confining background is possible as a solution of the
dilaton-gravity coupled equations in a deformed AdS model. It
means that our solution allows to obtain a spectrum of high-spin
mesons very close to Regge trajectories for the lower excited
states (where we have experimental data) and an exact linear Regge
trajectory for very high excitations. We solve self-consistently
the dilaton-gravity model, i.e., we adopt an active dilaton in
contrast to the Soft-Wall model where a passive dilation was
considered. Our model belongs to the general class of "Improved
AdS/QCD theories" proposed recently by G\"{u}rsoy, Kiritsis and
Nitti \cite{Kiritsis}.

\section{Dynamical AdS/QCD model}

In this section we will make a review of the Dynamical AdS/QCD
model as proposed by de Paula, Frederico, Forkel and Beyer
\cite{dePaulaPRD09}. Let's take the action for a five-dimensional
gravity coupled to a dilaton field:
\begin{eqnarray}
S =\int \frac{d^5x}{2\kappa^{2}} \sqrt{g} \left( -\emph{R} -
V(\Phi)
+\frac{1}{2}g^{MN}\partial_{M}\Phi\partial_{N}\Phi\right),
\nonumber \label{actiongd}
\end{eqnarray}
\noindent where $\kappa$ is the Newton constant in $5$ dimensions
and $V(\Phi)$ is the scalar-field potential . We will be
restricted to a metric family given by:
$$
g_{MN}=e^{-2A(z)}\eta_{MN}, $$  \noindent where $\eta_{MN}$ is the
Minkowski metric.

Minimizing the action we obtain the coupled Einstein equations
\begin{eqnarray}
6A'^{2} - \frac{1}{2}\Phi'^{2} + e^{-2A(z)} V( \Phi ) = 0, \label{einsteinzz}
\\
-3A'^{2} + 3A'' - \frac{1}{2} \Phi'^{2} - e^{-2A(z)} V(\Phi) = 0,
\label{einstein00} \\
\Phi '' - 3 A' \Phi '- e^{-2A(z)}\frac{dV}{d\Phi} = 0. \label{dilatonequation}
\end{eqnarray}
See that we can determine the dilaton field directly from the
metric model as: $$  \Phi'= \sqrt{3 A'^{2} + 3 A''}~,
\label{constrain}
$$
where we choose the positive sign for the root without loosing
generality. Substituting the dilaton field in equation
(\ref{einsteinzz}) we obtain the dilaton potential as
$$  V(\Phi) = \frac{3 e^{2A}}{2} \left(A''-3
A'^{2}\right), \label{vz} $$ by solving the dilaton-gravity 5d
Einstein equations.

\subsection{Hadronic Resonances}

As we have defined the dilaton-metric background of the model, we
are now able to calculate the meson mass spectrum in the spirit of
AdS/QCD duality. To do so, we utilize the AdS/CFT dictionary in
the sense that for each operator in the $4$d gauge theory there is
a field propagating in the bulk. For definiteness we follow the
notation of ref. \cite{Karch}. The $5$d action for a gauge field
$\phi_{M_{1}\dots M_{S}}$ of spin $S$ in the background is given
by
$$
I = \frac{1}{2} \int d^{5}x \sqrt{g} e^{-\Phi} \left(\nabla_{N}
\phi_{M_{1}\dots M_{S}} \nabla^{N} \phi^{M_{1}\dots M_{S}}
\right).
$$
As in \cite{Karch} and \cite{KatzLewandowski}, we utilize the
axial gauge. To this end, we introduce new spin fields
$\widetilde{\phi}_{\dots}= e^{2(S-1)A}\phi_{\dots}$. In terms of
this new field, the action is then given by
\begin{eqnarray}
I =\frac{1}{2} \int d^{5}x e^{-5A} e^{-\Phi} e^{-4(S-1)A}e^{2A(S+1)}\times \nonumber\\
\partial_{N}\widetilde{\phi}_{M_{1}\dots M_{S}}
\partial_{N}\widetilde{\phi}_{M_{1}\dots M_{S}} ~.
\label{eqn:newS}
\end{eqnarray}
Using (\ref{eqn:newS}) the equation for the modes
$\widetilde{\phi}_{n}$ of the higher spin field
$\widetilde{\phi}_{...}$ is derived, viz.
$$
\partial_{z}\left(e^{-B}\partial_{z}\tilde{\phi_{n}}\right)
+ m_{n}^{2}e^{-B}\tilde{\phi_{n}} = 0,
$$
where $B = A (2S-1) + \Phi$. Via the substitution
$\tilde{\phi_{n}}= e^{B/2}\psi_{n}$, one obtains a Sturm-Liouville
equation
$$
\left(-\partial_{z}^{2}+ {\mathcal V}_{eff}(z)\right)\psi_{n} = m_{n}^{2}\psi_{n},
$$
where the $B$ dependent term in this equation may be interpreted
as an effective potential for the string mode, written as
$$
{\mathcal V}_{eff}(z)=\frac{B'^{2}(z)}{4} -\frac{B''(z)}{2}.
$$
Hence, for each metric $A$ and dilaton field $\Phi$ we get
$$
{\mathcal V}_{eff}(z)= A'^{2}+\frac{5}{4}A''
 -\sqrt{3}~\frac{A''' +4A'A''+
2A'^3}{4\sqrt{A'^{2}+A''}}$$
$$ +
S^{2} A'^{2}+S\left(A'\sqrt{3A'^{2}+3A''}-A'^{2}-A''\right) ~ ,$$
consistent with the solutions of the Einstein equations. By
solving the eigenvalue mode equation, we obtain a mass spectrum
$m_n^2$ starting from the effective potential. Due to the
gauge/gravity duality this mass spectrum corresponds to the
mesonic resonances in the $4$d space-time.

 \subsection{Scalar Mesons}

For scalars we can repeat the same derivation as we did in the
last section except for the fact that for these mesons obviously
we do not have to make any gauge choice. In particular scalar
mesons were analyzed in \cite{Schmidt_Scalar,Colangelo_Scalar}.
However, both works do not include the sigma.

We start from the action \cite{dePaulaArXiv09} \small{
$$
I =\int \frac{d^4x}{2} dz\sqrt{\left\vert g\right\vert }
\left(g^{\mu\nu}\partial_{\mu}
\varphi(x,z)\partial_{\nu}\varphi(x,z) - m_5^2\varphi^{2} \right),
$$}
where $m_5^2=M_{5}^{2}/\Lambda_{QCD}^{2}$, that describes a scalar
mode propagating in the dilaton-gravity background. We factorize
in terms of the holographic coordinate
$\varphi(x,z)=e^{iP_{\mu}x^{\mu}}\varphi(z)$,
$P_{\mu}P^{\mu}=m^{2}$.

The string modes of the massive scalar field $\varphi$ can be
rewritten in terms of the reduced amplitudes $\psi
_{n}(z)=\varphi_{n}(z) \times e^{-(3A+\Phi)/2}$ which satisfy the
Sturm-Liouville equation, as we have written before for the spin
states. The scalar string-mode potential given by
\begin{equation} ~~~~~~~\mathcal{V}(z)=\frac{B^{\prime
}{}^{2}(z)}{4}-\frac{B^{\prime \prime
}(z)}{2}+\frac{M_{5}^{2}}{\Lambda_{QCD}^{2}}e^{-2A(z)}, \label{vs}
\end{equation}%
with $B=3A+\Phi$.  The gauge/gravity dictionary identifies the
eigenvalues $m_{n,S}^{2}$ with the squared meson mass spectrum of
the boundary gauge theory.

The AdS/CFT correspondence states that the wave function should
behave as $z^{\tau}$, where $\tau = \Delta - \sigma$ (conformal
dimension minus spin) is the twist dimension for the corresponding
interpolating operator that creates the given state configuration
\cite{Polchinski}. The five-dimensional mass chosen as
\cite{Witten98} $M_{5}^{2}=\tau(\tau-4),$ fixes the UV limit of
the dual string amplitude with the twist dimension.

\section{Analytical Analysis}

In order to present an analytical view of confinement from the
gravity model, as have been done in refs.
\cite{Kiritsis,dePaulaPRD09}, we will focus on a very simple
polynomial metric, representing an IR deformation of AdS metric:
\begin{equation}
~~~~~~~~~~~~~~A(z) = \log \left(z \right)+z^{\lambda}+\dots ~,
\label{polinomial_metric}
\end{equation}
where $\lambda$ is a real parameter. Our units are such that the
$\mathrm{AdS}_{5}$ radius is unity. The first term reflects the
AdS metric that dominates the UV limit. The second term is the
leading one in the IR region and any subleading term is irrelevant
for the present discussion of the Regge trajectory for the high
excited string states dual to mesons. The dilaton field is
obtained by integrating Eq. (\ref{constrain}) with the boundary
condition $\Phi(0)=0$. In particular one gets, for $z\rightarrow
0$ and $z\rightarrow \infty$, respectively $$
 \Phi(z)\sim c_0
z^{\frac{\lambda}{2}} ~~and~~ \Phi(z)\sim c_\infty z^\lambda ,
\label{phiz}
$$
where $c_0=2\sqrt{3 \left(1+1/ \lambda\right)}$ and $c_\infty=\sqrt{3}$.

The dilaton potential for $z\rightarrow 0$ is
$$
V(\Phi)\sim-6 +\frac{3}{2 c_0^2}(\lambda+1)(\lambda-8)\Phi^2
 \ ,
$$
and for $z\rightarrow \infty$ we obtain:
$$
V(\Phi)\sim-\frac{9}{2 c_\infty^2} \lambda^2\Phi^2 e^{2\Phi/c_\infty} \ ,
$$
which diverges exponentially reflecting the exponential form of
the metric model. As an example, let us discuss the  UV and IR
properties of the effective potential for nonzero spin mesons. For
small values of $z$ it can be easily expanded giving:
\begin{eqnarray}
{\mathcal V}_{eff}(z) = \frac{S^2-\frac14}{z^2}+\sqrt{3(\lambda^2+\lambda)}\left(S-\frac{\lambda}{4}\right)z^{\frac{\lambda}{2}-2} \nonumber \\
+\frac{\lambda}{4}\left(8S^2-S(4\lambda+4)+5\lambda+3\right)z^{\lambda-2}+ \dots , \label{eqn:UVeffP}
\end{eqnarray}
which shows a spin dependence in all lower order terms. In the IR
limit the metric (\ref{polinomial_metric}) leads to the following
effective potential
\begin{equation}
~~~~~~~~~~{\mathcal V}_{eff}(z) \rightarrow \frac{ \lambda^2}{4}
(2S-1+\sqrt{3})^{2}\; z^{2\lambda-2}, \label{eqn:effP}
\end{equation}
which presents a discrete spectrum for the normalizable string
modes if $\lambda > 1$. It is worthwhile to point out that the
analysis of the effective potential gives a constraint for a
confining metric consistent with the one found in the analysis of
the Wilson loop (see \cite{Kiritsis}).

\section{Phenomenological Results}

\begin{figure}[tbh]
\centerline{\epsfig{figure=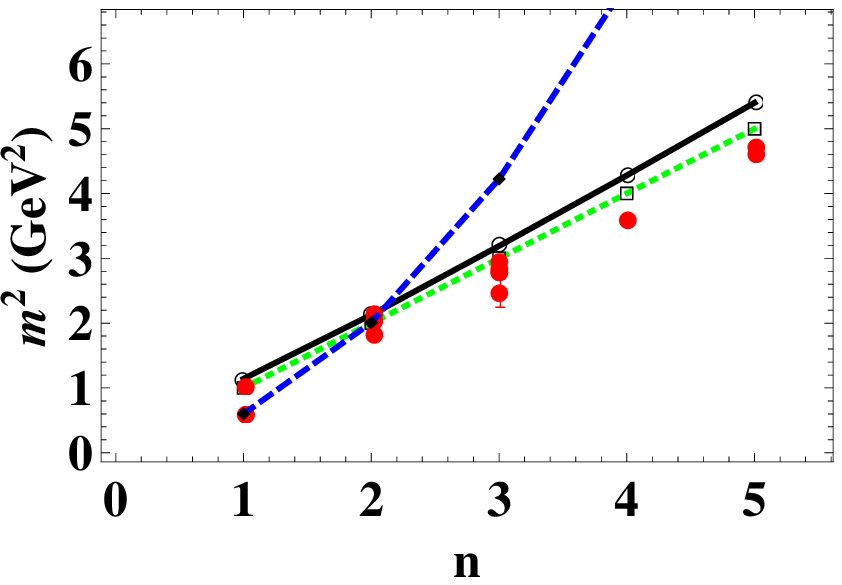,width=4.cm,height=3.8cm}
\epsfig{figure=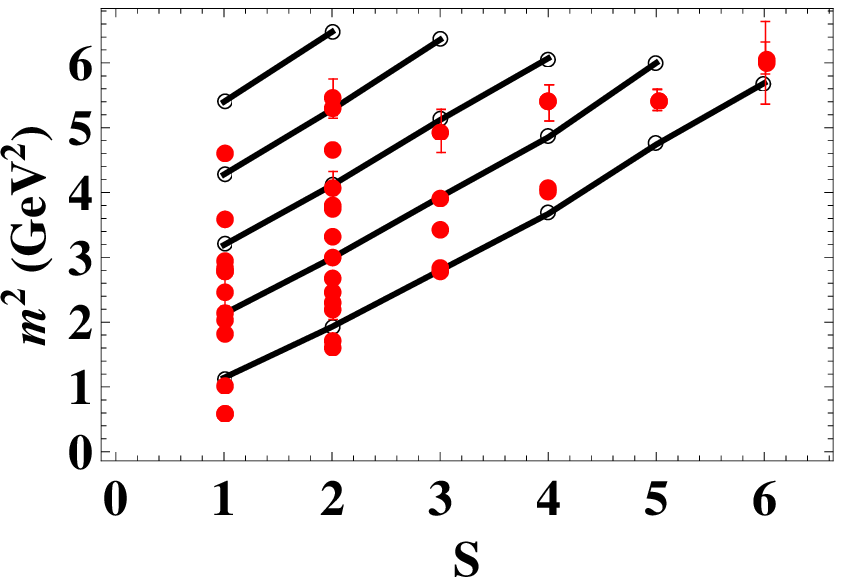,width=4.cm,height=3.8cm}}
\caption{Radial excitations of the rho meson in the hard-wall
(dashed line), soft-wall \protect\cite{Karch} (dotted line) and
our dynamical soft-wall (solid line, for $\Lambda _{QCD}=0.3$ GeV)
backgrounds(left panel). Dynamical AdS/QCD squared mass
predictions of spin excitations (right panel). Experimental data
from \cite{pdg}. } \label{Fig1}
\end{figure}

\begin{figure}[hbt]
\centerline{\epsfig{figure=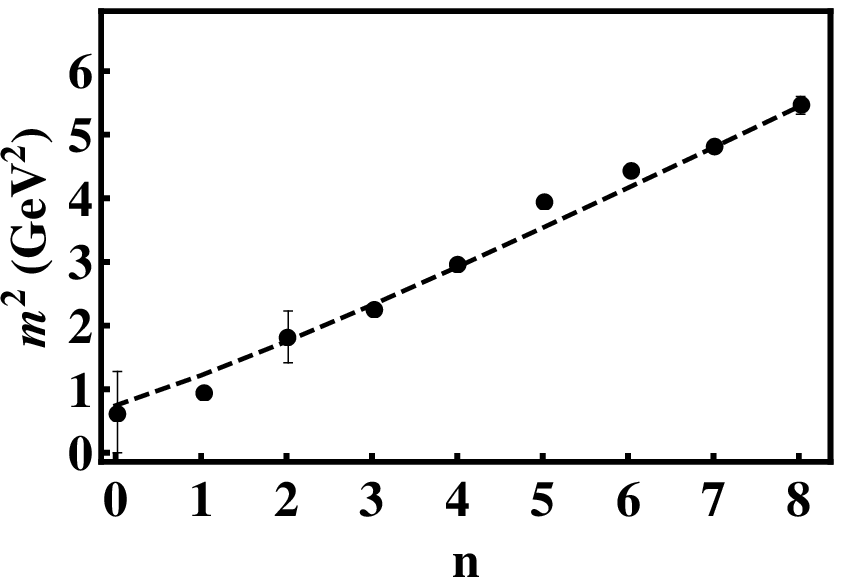,width=4.cm,height=3.8cm}
\epsfig{figure=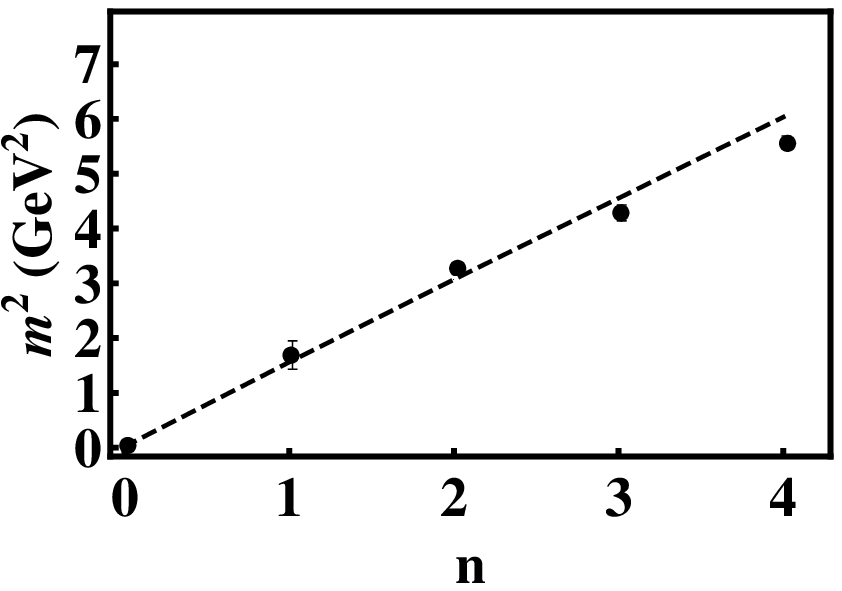,width=4.cm,height=3.8cm}} \caption{Regge
trajectory for $f_0$ (left panel) and pion (right panel) from  the
Dynamical AdS/QCD model with $\Lambda _{QCD}=0.3$ GeV.
Experimental data from \cite{pdg}. } \label{Fig2}
\end{figure}

In our derivations we reduced the problem of modelling hadronic
resonances by solving a Sturm-Liouville equation with a given
potential (\ref{vs} or \ref{eqn:effP}). The amazing point is that
the effective potential depends only on the metric, which
automatically constructs a self consistent dilaton-gravity
background. Therefore our modelling is at the level of proposing a
metric ansatz that generates the experimental data available
(mass, decay constants, form factors,...). In addition this metric
has to satisfy the following conditions: i) in the UV it has to
become AdS, because QCCD is conformal in this limit and we have to
recover the Maldacena duality; ii) in the IR the warp factor has
to have a polynomial power of $\lambda=2$, in order to have
confinement by the Wilson Loop criteria and linear Regge
trajectories. Our ansatz is:
$$
A(z)= Log(\xi z \Lambda_{QCD}) + \frac{(\xi z\Lambda
_{QCD})^{2}}{1+e^{(1-\xi z\Lambda _{QCD})}}\ , \label{cnew}
$$
where $\xi$ is a scale transformation that connects the gravity
background in which different string modes dual propagate in the
holographic coordinate. For scalars $\xi = 0.58$. To distinguish
the pion states in our model, the fifth dimensional mass was
rescaled according to $M_{5}^{2} \rightarrow M_{5}^{2}+\lambda
z^{2}$ (see \cite{FBT}). The constraints are the pion mass, the
slope of the Regge trajectory and the twist 2 from the operator
$\bar q \gamma^5 q$. The results for the pion Regge trajectory are
shown in figure 2 for $\xi=$0.88 and $\lambda=-2.19$GeV$^2$. For
high-spin mesons we have an equation to obtain the scale factor
$\xi=S^{-0.3329}$ (in particular, see that in our previous work
\cite{dePaulaPRD09} we also have the Regge trajectories with a
different metric ansatz). With the present metric ansatz we obtain
the Regge trajectories in agreement to experimental data for
vector, high-spin, scalar and pseudoscalar mesons.

\begin{table*}[htb]
\caption{Two-pion decay width and masses for the $f_{0}$ family.
Experimental values from Particle Data Group \cite{pdg}.
$^\dagger$Mixing angle of $20^o$. $^*$Fitted value.}
\label{table:width}
\newcommand{\m}{\hphantom{$-$}}
\newcommand{\cc}[1]{\multicolumn{1}{c}{#1}}
\renewcommand{\tabcolsep}{2pc} 
\renewcommand{\arraystretch}{1.2} 
\begin{tabular}{@{}lllll}
\hline
Meson                     & \cc{$M_{exp}$(GeV)} & \cc{$M_{th}$(GeV)} & \cc{$\Gamma^{exp}_{\pi\pi}$(MeV)} & \cc{$\Gamma^{th}_{\pi\pi}$(MeV)} \\
\hline
$f_{0}(600) $ & 0.4 - 1.2        & 0.86   & 600 - 1000    & 535 \\
$f_{0}(980) $ & 0.98$\pm$ 0.01   & 1.10   & $\sim$15-80   & 42$^\dagger$  \\
$f_{0}(1370)$ & 1.2 - 1.5        & 1.32   & $\sim$41-141  & 141   \\
$f_{0}(1500)$ & 1.505$\pm$0.006  & 1.52   & 38$\pm$3      & 38$^*$  \\
$f_{0}(1710)$ & 1.720$\pm$0.006  & 1.70   & $\sim$ 0-6    & 5  \\
$f_{0}(2020)$ & 1.992$\pm$0.016  & 1.88   &  ---          & 0.0  \\
$f_{0}(2100)$ & 2.103$\pm$0.008  & 2.04   &  ---          & 1.2  \\
$f_{0}(2200)$ & 2.189$\pm$0.013  & 2.19   & ---           & 2.5  \\
$f_{0}(2330)$ & 2.29-2.35        & 2.33   &  ---          & 2.8  \\
\hline
\end{tabular}\\[2pt]
\end{table*}

\section{Decay Amplitudes}

The $f_0$'s partial decay width into $\pi\pi$ are calculated from
the overlap integral of the normalized string amplitudes
(Sturm-Liouville form) in the holographic coordinate  dual to the
scalars $(\psi_n)$ and pion $(\psi_\pi)$ states,
\begin{equation}
~~~~~~~~~~~~~~~h_{n} =k\int^\infty_0 dz ~\psi_\pi^2(z)\psi_n(z) \
, \label{dwover}
\end{equation}
where $k$ is a constant with dimension $\sqrt{mass}$ fitted to the
experimental value of the $f_0(1500)\to \pi\pi$ partial decay
width. The overlap integral is our guess for the dual
representation of the transition amplitude $S\to PP$ and therefore
the decay width is given by $\Gamma_{\pi \pi}^{n} =
\frac{1}{8\pi}|h_{n}|^{2}\frac{p_\pi}{m_n^2}\ ,$ where $p_\pi$ is
the pion momentum in the meson rest frame. The Sturm-Liouville
amplitudes of the scalar (pseudoscalar) modes are normalized just
as a bound state wave function in quantum mechanics
\cite{RadyushkinPRD2007,BrodskyPRD2008025}, which also corresponds
to a normalization of the string amplitude
$$
\int_{0}^{\infty}dz\psi_{m}(z)\psi_{n}(z)=\delta_{mn} \ . \label{norm}
$$

The two-pion partial decay width for the $f_0$'s present in the
particle listing of PDG, are calculated with Eq. (\ref{dwover})
and shown in Table I. The width of $f_0(1500)$ is used as
normalization for the parameter . In particular for $f_0(600)$ the
model gives a width of about 500 MeV, while its mass is 860 MeV. A
large range of experimental values is quoted in PDG for the sigma
mass and width (see Table I).

The E791 experiment quotes $m_\sigma=$ 478$^{+24}_{-23}\pm 17$ MeV
and $\Gamma_\sigma=324^{+42}_{-40}\pm 21$ MeV \cite{Aitala_sigma},
which in has a width consistent with our model while the mass
appears somewhat larger. The CLEO collaboration \cite{CLEO2002}
quotes $m_\sigma=$ 513$\pm 32$ MeV and $\Gamma_\sigma=335\pm 67$
MeV, and a recent analysis of the sigma pole in the $\pi\pi$
scattering amplitude from ref.\cite{Caprini06} gives $m_\sigma=$
441$^{+16}_{-8}$ MeV and $\Gamma_\sigma=544^{+18}_{-25}$ MeV.
Other analysis of the $\sigma$-pole in the $\pi\pi \to \pi\pi$
scattering amplitude present in the decay of heavy mesons
indicates a mass around 500 MeV \cite{bugg06}.

\section{Conclusions}

In this paper we presented a Dynamical AdS/QCD model applied to
light meson spectroscopy. We solved the dilaton-gravity coupled
equations within a metric model and obtained a linear confining
background for the string modes dual to mesons. We obtained a
spectrum of the light-favored high-spin, scalar and pseudoscalar
mesons in agreement with the experimental data. In addition we
calculated the decay amplitude for the $f_0$'s  into two pions to
further check the consistency of the physical scales of the model,
as this quantity is strongly sensitive to the relative size of the
different mesons.

In particular, the $f_0(980)$ is  identified with the first
excitation of the string model dual to $q\bar q$  state (see Table
I). We interpret this  shift to a value above the experimental
value, i.e., 1.1 GeV compared to 0.98 GeV as due to a  rescaling
of the string mass as in the pion case, that also should be the
case for the sigma. By increasing the excitation of the scalar
meson this shift tends to decrease (see $f_0(1500)$ in Table I).
We observed that the experimental value of $\Gamma_{\pi\pi}$ for
$f_0(980)$ is too small compared to our result. This indicated
that a strong mixing, e.g., of $s\bar{s}$ with light non-strange
quarks\cite{Bediaga}, should be present in the model. To account
for that a mixing angle for $f_0(980)$ of $\pm 20^o$ was obtained
from the partial width, and values between $\sim 12^{\circ}$  to
$28^{\circ}$ fits $\Gamma^{\pi\pi}$ within the experimental range.

Let us remind an interesting observation, that the string mode
amplitude could be identified with the valence light-front wave
function as pointed out by Brodsky and de T\'eramond
\cite{BrodskyPRL09}. The wave equation in the Sturm-Liouville form
is identified with the squared mass operator eigenvalue equation
for the valence component of the meson light-front wave function
(see also \cite{Schmidt_wavefunction}). An alternative way to
reproduce Regge trajectories using the anomalous dimension of the
operators are given by Vega and Schmidt \cite{Schmidt_Hadrons}.

As a next step, we are currently considering  the strange meson
sector\cite{AdSQCDStrange} within the Dynamical AdS/QCD model. For
a future challenge we also want to introduce finite temperature
and calculate the spectrum as done in ref. \cite{Miranda} and
compare to recent Lattice results\cite{panero}.

We acknowledge partial support from CAPES, FAPESP and CNPq.

\end{document}